\documentclass[useAMS,usenatbib]{mn2e}
\usepackage{graphicx}
\usepackage{longtable}
\usepackage{lscape}
\newcommand\nion[2]{#1\,\lowercase{{\sc #2}}}

\def\kmsec{\mbox{km~s$^{\rm -1}$}}

\def\teff{T_{\rm eff}}
\def\eteff{\sigma_{\teff}}
\def\tlte{T_{\rm LTE\mbox{-}Fe}}

\def\tfin{T_{\rm Opt}}

\def\tr11{T_{\rm R11}}
\def\etr11{\sigma_{\tr11}}
\def\tbal{T_{\rm Bal}}
\def\etbal{\sigma_{\tbal}}
\def\tirfm{T_{\rm IRFM}}
\def\etirfm{\sigma_{\tirfm}}
\def\tint{T_{\rm Int}}
\def\logg{\log~g}
\def\logglte{\log~g_{\rm LTE\mbox{-}Fe}}
\def\loggpi{\log~g_{\pi}}
\def\loggf{\log~g_{\rm NLTE\mbox{-}Opt}}

\def\vt{v}
\def\vtlte{v_{\rm LTE\mbox{-}Fe}}
\def\vtf{v_{\rm NLTE\mbox{-}Opt}}
\def\fehlte{\rm[Fe/H]_{\rm LTE\mbox{-}Fe}}
\def\fehf{\rm[Fe/H]_{\rm NLTE\mbox{-}Opt}}
\def\feh{\rm[Fe/H]}

\def\jks{J\mbox{-}K_{\rm S}}

\def\simgt{\lower.5ex\hbox{$\; \buildrel > \over \sim \;$}}
\def\simlt{\lower.5ex\hbox{$\; \buildrel < \over \sim \;$}}

\title[A novel approach to determining stellar parameters]{Unveiling systematic biases in 1D LTE excitation-ionisation balance of Fe for FGK stars. A novel approach to determination of stellar parameters.}

\author[Ruchti et al.]
{Gregory~R.~Ruchti,$^{1}$\thanks{Current address: Lund Observatory, Department of Astronomy and Theoretical Physics, Box 43, SE-22100, Lund, Sweden: greg@astro.lu.se} 
Maria~Bergemann,$^{1}$ 
Aldo~Serenelli,$^{2}$
\newauthor
Luca~Casagrande$^{3}$ 
and Karin~Lind$^{1}$ \\
$^{1}$Max Planck Institut f\"ur Astrophysik, Karl-Schwarzschild-Str. 1, 85741 Garching, Germany \\
$^{2}$Instituto de Ciencias del Espacio (CSIC-IEEC), Facultad de Ciencias, Campus UAB, 08193 Bellaterra, Spain\\
$^{3}$Research School of Astronomy \& Astrophysics, Mount Stromlo Observatory, The Australian National University, ACT 2611, Australia\\
}

\date{Accepted 2012 October 30.  Received 2012 September 26}

\pagerange{\pageref{firstpage}--\pageref{lastpage}} \pubyear{2012}

\begin{document}

\label{firstpage}

\maketitle

\begin{abstract}

We present a comprehensive analysis of different techniques available for the spectroscopic analysis of FGK stars, and provide a recommended methodology which efficiently estimates accurate stellar atmospheric parameters for large samples of stars.  Our analysis includes a simultaneous equivalent width analysis of \nion{Fe}{I} and \nion{Fe}{II} spectral lines, and for the first time, utilises on-the-fly NLTE corrections of individual \nion{Fe}{I} lines.  We further investigate several temperature scales, finding that estimates from Balmer line measurements provide the most accurate effective temperatures at all metallicites.  We apply our analysis to a large sample of both dwarf and giant stars selected from the RAVE survey.  We then show that the difference between parameters determined by our method and that by standard 1D LTE excitation-ionisation balance of Fe reveals substantial systematic biases: up to $400$~K in effective temperature, $1.0$~dex in surface gravity, and $0.4$~dex in metallicity for stars with $\feh\sim-2.5$.  This has large implications for the study of the stellar populations in the Milky Way.

\end{abstract}


\section{Introduction}
\label{sec-intro}

The fundamental atmospheric (effective temperature, surface gravity, and metallicity) and physical (mass and age) parameters of stars provide the major observational foundation for chemo-dynamical studies of the Milky Way and other galaxies in the Local Group.  With the dawn of large spectroscopic surveys to study individual stars, such as SEGUE \citep{yanny09}, RAVE \citep{steinmetz06}, Gaia-ESO \citep{gilmore12}, and HERMES \citep{barden08}, these parameters are used to infer the characteristics of different populations of stars that comprise the Milky Way.

Stellar parameters determined by spectroscopic methods are of a key importance. The only way to accurately measure metallicity is through spectroscopy, which thus underlies photometric calibrations \citep[e.g.,][]{holmberg07,an09,arnadottir10,casagrande11}, while high-resolution spectroscopy is also used to correct the low-resolution results \citep[e.g.,][]{carollo10}. The atmospheric parameters can all be estimated from a spectrum in a consistent and efficient way.  This also avoids the problem of reddening inherent in photometry since spectroscopic parameters are not sensitive to reddening.  The spectroscopic parameters can then be used alone or in combination with photometric information to fit individual stars to theoretical isochrones or evolutionary tracks to determine the stellar mass, age, and distance of a star. 

A common method for deriving the spectroscopic atmospheric parameters is to use the information from \nion{Fe}{I} and \nion{Fe}{II} absorption lines under the assumption of hydrostatic equilibrium (HE) and local thermodynamic equilibrium (LTE).  Many previous studies have used some variation of this technique (e.g., ionisation or excitation equilibrium) to determine the stellar atmospheric parameters and abundances, and henceforth distances and kinematics, of FGK stars in the Milky Way.  For example, some have used this procedure to estimate the effective temperature, surface gravity, and metallicity of a star \citep[e.g.,][]{fulbright00,prochaska00,johnson02}, while others use photometric estimates of effective temperature in combination with the ionisation equilibrium of the abundance of iron in LTE to estimate surface gravity and metallicity \citep[e.g.,][]{mcwilliam95,francois03,bai04,allendep06,lai08}.

However, both observational \citep[e.g.,][]{fuhrmann98,ivans01,ruchti11,bruntt12} and theoretical evidence \citep[e.g.,][]{thevenin99,asplund05,mashonkina11} suggest that systematic biases are present within such analyses due to the breakdown of the assumption of LTE.   More recently, \citet{bergemann12} and \citet{lind12} quantified the effects of non-local thermodynamic equilibrium (NLTE) on the determination of surface gravity and metallicity, revealing very substantial systematic biases in the estimates at low metallicity and/or surface gravity.  It is therefore extremely important to develop sophisticated methods, which reconcile these effects in order to derive accurate spectroscopic parameters.

This is the first in a series of papers, in which we develop new, robust methods to determine the fundamental parameters of FGK stars and then apply these techniques to large stellar samples to study the chemical and dynamical properties of the different stellar populations of the Milky Way.  In this work, we utilise the sample of stars selected from the RAVE survey originally published in \citet[][hereafter R11]{ruchti11} to formulate the methodology to derive very accurate atmospheric parameters.  We consider several temperature scales and show that the Balmer line method is the most reliable among the different methods presently available.  Further, we have developed the necessary tools to apply on-the-fly NLTE corrections\footnote{``NLTE correction" refers to the difference between the abundance of iron computed in LTE and NLTE obtained from a line with a given equivalent width.}  to \nion{Fe}{I} lines, utilising the grid described in \citet{lind12}.  We verify our method using a sample of standard stars with interferometric estimates of effective temperature and/or {\it Hipparcos} parallaxes.  We then perform a comprehensive comparison to standard 1D, LTE techniques for the spectral analysis of stars, finding significant systematic biases.

\section{Sample Selection and Observations}

NLTE effects in iron are most prominent in low-metallicity stars \citep{lind12,bergemann12}.  We therefore chose the metal-poor sample from R11 for our study.  These stars were originally selected for high-resolution observations based on data obtained by the RAVE survey in order to study the metal-poor thick disk of the Milky Way.  Spectral data for these stars were obtained using high-resolution echelle spectrographs at several facilities around the world.  

Full details of the observations and data reduction of the spectra can be found in R11.  Briefly, all spectrographs delivered a resolving power greater than 30,000 and covered the full optical wavelength range.  Further, nearly all spectra had signal-to-noise ratios greater than $100:1$~per~pixel.  The equivalent widths (EWs) of both \nion{Fe}{I} and \nion{Fe}{II} lines, taken from the line lists of \citet{fulbright00} and \citet{johnson02}, were measured using the ARES code \citep{sousa07}.  However, during measurement quality checks, we found that the continuum was poorly estimated for some lines.  We therefore determined EWs for these affected lines using hand measurements.  

\section{Stellar Parameter Analyses}
\label{sec-par}

We computed the stellar parameters for each star using two different methods.  

In the first method, which is commonly used in the literature, we derived an effective temperature, $\tlte$, surface gravity, $\logglte$, metallicity, $\fehlte$, and microturbulence, $\vtlte$, from the ionisation and excitation equilibrium of Fe in LTE.   This is hereafter denoted as the LTE-Fe method.  We used an iterative procedure that utilised the \texttt{MOOG} analysis program \citep{sneden73} and 1D, plane-parallel \texttt{ATLAS-ODF} model atmospheres from Kurucz\footnote{See http://kurucz.harvard.edu/.} computed under the assumption of LTE and HE.  In our procedure, the stellar effective temperature was set by minimising the magnitude of the slope of the relationship between the abundance of iron from \nion{Fe}{I} lines and the excitation potential of each line.  Similarly, the microturbulent velocity was found by minimising the slope between the abundance of iron from \nion{Fe}{I} lines and the reduced EW of each line.  The surface gravity was then estimated by minimising the difference between the abundance of iron measured from \nion{Fe}{I} and \nion{Fe}{II} lines.  Iterations continued until all of the criteria above were satisfied.  Finally, $\fehlte$ was chosen to equal the abundance of iron from the analysis.  Our results for this method are described in Section~\ref{sec-lte}.

The second method, denoted as the NLTE-Opt method, consists of two parts.  First, we determined the optimal effective temperature estimate, $\tfin$, for each star (see Section~\ref{sec-temp} for more details).  Then, we utilised \texttt{MOOG} to compute a new surface gravity, $\loggf$, metallicity, $\fehf$, and microturbulence, $\vtf$.  This was done using the same iterative techniques as the LTE-Fe method, that is the ionisation balance of the abundance of iron from \nion{Fe}{I} and \nion{Fe}{II} lines.  

There are, however, three important differences.  First, the stellar effective temperature was held fixed to the optimal value, $\tfin$.  Second, we restricted the analysis to Fe lines with excitation potentials above 2~eV, since these lines are less sensitive to 3D effects as compared to the low-excitation lines \citep[see the discussion in][]{bergemann12}.  Third, the abundance of iron from each \nion{Fe}{I} line was adjusted according to the NLTE correction for that line at the stellar parameters of the current iteration in the procedure.  The NLTE corrections were determined using the NLTE grid computed in \citet{lind12} and applied on-the-fly via a wrapper program to \texttt{MOOG}.  Note that the NLTE calculations presented in \citet{lind12} were analogously calibrated using the ionisation equilibria of a handful of well-known stars.  Our extended sample, including more stars with direct measurements of surface gravity and effective temperature, provide support for the realism of this calibration.  The grid extends down to $\logg=1$.  We imposed a routine which linearly extrapolated the NLTE corrections to below this value. The results of extrapolations were checked against NLTE grids presented in \citet{bergemann12a} and no significant differences were found.  Further, \citet{lind12} found very small NLTE corrections for \nion{Fe}{II} lines.  We therefore do not apply any correction to the \nion{Fe}{II} lines.

Iterations continued until the difference between the average abundance of iron from the \nion{Fe}{II} lines and the NLTE-adjusted \nion{Fe}{I} lines were in agreement (within $\pm0.05$~dex) and the slope of the relationship between the reduced EW of the \nion{Fe}{I} lines and their NLTE-adjusted iron abundance was minimised.  Sections~\ref{sec-temp} and \ref{sec-nlte} describe our final stellar parameter estimates for this method.

\section{Initial LTE-Fe Parameters}
\label{sec-lte}

The initial LTE-Fe stellar parameters for our sample stars are listed in Table~\ref{tab-par}.  Residuals in the minimizations of this technique gave typical internal errors of 0.1~dex in both $\logglte$ and $\fehlte$ and $\sim55$~K in $\tlte$.  As we show in the following sections, these small internal errors can be quite misleading as they are not representative of the actual accuracy of stellar parameter estimates.  Often, especially in metal-poor stars, estimates of $\teff$, $\logg$, and $\feh$, that result from this method are far too low when compared to other, more accurate data (cf., R11).

\begin{table*}
\centering
\setlength{\tabcolsep}{0.05in}
 \begin{minipage}{200mm}
  \caption{Atmospheric Stellar Parameter Data}
  \label{tab-par}
  \begin{tabular}{@{}rrrrrrrrrrrrrrrrr@{}}
  \hline
  & \multicolumn{5}{c}{LTE-Fe} & & & & & & & \multicolumn{5}{c}{NLTE-Opt} \\
  \cline{2-6} \cline{13-17} \\
   Star & $\teff$ & $\eteff$ & $\logg$ & $\feh$ & $\vt$ & $E(B\mbox{-}V)$ & $\tirfm$ & $\etirfm$ & $\tr11$ & $\tbal$ & $\etbal$ & $\teff$ & $\eteff$ & $\logg$ & $\feh$ & $\vt$ \\
          & (K) & (K) & ($\pm0.1$) & ($\pm0.1$)  & & & (K) & (K) & ($\pm140$~K) & (K) & (K) & (K) & (K) & ($\pm0.1$) & ($\pm0.1$) & \\
          \hline
C0023306-163143 & 5128 & 58 & 2.40 & -2.63 & 1.3 & -- & -- & -- & 5528 & 5400 & 100 & 5443 & 101 & 3.20 & -2.29 & 0.9 \\
C0315358-094743 & 4628 & 40 & 1.51 & -1.40 & 1.5 & -- & -- & -- & 4722 & 4800 & 100 & 4774 & 89 & 2.06 & -1.31 & 1.6 \\
C0408404-462531 & 4466 & 40 & 0.25 & -2.25 & 2.2 & -- & -- & -- & 4600 & -- & -- & 4600 & 90 & 1.03 & -2.10 & 2.1 \\
C0549576-334007 & 5151 & 50 & 2.53 & -1.94 & 1.3 & -- & -- & -- & 5379 & 5400 & 100 & 5393 & 82 & 3.16 & -1.70 & 1.1 \\
C1141088-453528 & 4439 & 40 & 0.39 & -2.42 & 2.1 & -- & -- & -- & 4592 & 4500 & 200 & 4562 & 123 & 1.10 & -2.28 & 1.9 \\
\hline	
\end{tabular}
\end{minipage}
This table is published in its entirety in the electronic edition of the MNRAS. A portion is shown here for guidance regarding its form and content.
\end{table*}

\section{Effective Temperature Optimisation}
\label{sec-temp}

It was found in \citet{bergemann12} and \citet{lind12} that taking into account NLTE in the solution of excitation equilibrium does not lead to a significant improvement of the stellar effective temperature. This was also supported by our test calculations for a sub-sample of stars.  \nion{Fe}{I} lines formed in LTE or NLTE are still affected by convective surface inhomogeneities and overall different mean temperature/density stratifications, which are most prominent in strong low-excitation \nion{Fe}{I} lines \citep{shchukina05,bergemann12}. Using 1D hydrostatic models with either LTE or NLTE radiative transfer thus leads to effective temperature estimates that are too low when the excitation balance of \nion{Fe}{I} lines is used (see below). It is therefore important that the stellar effective temperature be estimated by other means.

\subsection{Three Effective Temperature Scales}

We used three different methods to compute the effective temperature.  

\begin{figure}
\includegraphics[width=84mm]{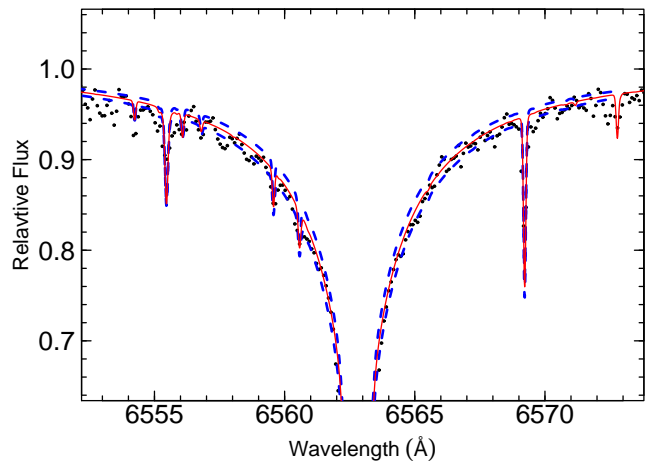}
\caption{Example fit to H$_{\alpha}$ in the spectrum of the metal-poor ($\feh\sim-1.0$) dwarf, J142911.4-053131.  The solid, red curve shows the best fit to the data (at 5700~K), while the blue, dashed curves represent $\pm100$~K around the fit.}
\label{fig-bal3}
\end{figure}

\begin{figure}
\includegraphics[width=84mm]{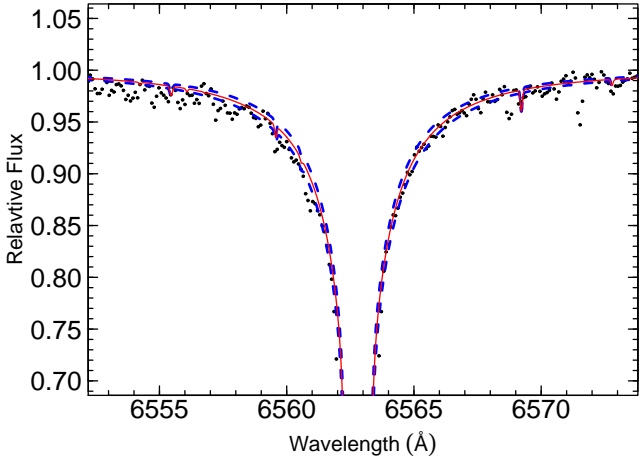}
\caption{Example fit to H$_{\alpha}$ in the spectrum of the metal-poor ($\feh\sim-2.3$) giant, J230222.8-683323.  The solid, red curve shows the best fit to the data (at 5200~K), while the blue, dashed curves represent $\pm100$~K around the fit.}
\label{fig-bal5}
\end{figure}

The first estimate, $\tbal$, was derived from the wings of the Balmer lines, which is among the most reliable methods available for the effective temperature determination of FGK stars \citep[e.g.,][]{fuhrmann93,fuhrmann98,barklem02,cowley02,gehren06,mashonkina08}.  The only restriction of this method is that for stars cooler than $4500$ K, the wings of \nion{H}{I} lines become too weak to allow reliable determination of $\teff$. Profile fits of H$_\alpha$ and H$_\beta$ lines were performed by careful visual inspection of different portions of the observed spectrum in the near and far wings of the Balmer lines which were free of contaminant stellar lines.  Figures~\ref{fig-bal3} and \ref{fig-bal5} show two example fits to H$_{\alpha}$.  Note that the Balmer lines were self-contained within a single order in each spectrum.  Therefore, we did not use neighbouring orders for the continuum normalisation.

Theoretical profiles were computed using the SIU code with \texttt{MAFAGS-ODF} model atmospheres \citep{fuhrmann98,grupp04a}.  Same as \texttt{ATLAS-ODF} (Section \ref{sec-par}), the MAFAGS models were computed with Kurucz opacity distribution functions, thus the differences between the model atmosphere stratifications are expected to be minimal in our range of stellar parameters. For self-broadening of H lines, we used the \citet{ali65} theory. As shown by \citet{grupp04a} this method successfully reproduces the Balmer line spectrum of the Sun within $20$ K, and provides accurate stellar parameters that agree very well with {\it Hipparcos} astrometry \citep{grupp04b}. The errors are obtained directly from profile fitting, and they are largely internal, $\pm50$ to 100~K.

A key advantage of the Balmer lines is that they are insensitive to interstellar reddening, which affects photometric techniques (see below).  However, the Balmer line effective temperature scale could be affected by systematic biases, caused by the physical limitation of the models. The influence of deviations from LTE in the \nion{H}{I} line formation in application to cool metal-poor stars was studied by \citet{mashonkina08}. Comparing our results to the NLTE estimates by \citet{mashonkina08} for the stars in common, we obtain: $\Delta \teff$(our - M08) $= -70$ K (HD 122563 metal-poor giant), $\Delta \teff$(our - M08) $= 50$ K (HD 84937, metal-poor turn-off). The difference is clearly within the $\teff$ uncertainties. On the other side, it should be kept in mind that the atomic data for NLTE calculations for hydrogen are of insufficient quality and, at present, do not allow accurate quantitative assessment of NLTE effects in H, as elaborately discussed by \citet{barklem07}. Likewise, the influence of granulation is difficult to assess. \citet{ludwig09} presented 3D effective temperature corrections for Balmer lines for a few points on the HRD, for which 3D radiative-hydrodynamics simulations of stellar convection are available. For the Sun\footnote{Here, we use their results obtained with $\alpha_{\rm MLT} = 0.5$ consistent with the \texttt{MAFAGS-ODF} model atmospheres adopted here.}, they find $\Delta \teff (\rm{3D - 1D}) \approx 35$ K, and for a typical metal-poor subdwarf with [Fe/H] $=-2$, $\Delta \teff (\rm{3D - 1D})$ of the order $50$ to $80$ K (average over H$_\alpha$, H$_\beta$, and H$_\gamma$). However, in the absence of consistent 3D NLTE calculations, it is not possible to tell whether 3D and NLTE effects will amplify or cancel for FGK stars. Thus, we do not apply any theoretical corrections to our Balmer effective temperatures.

\begin{figure}
\includegraphics[width=84mm]{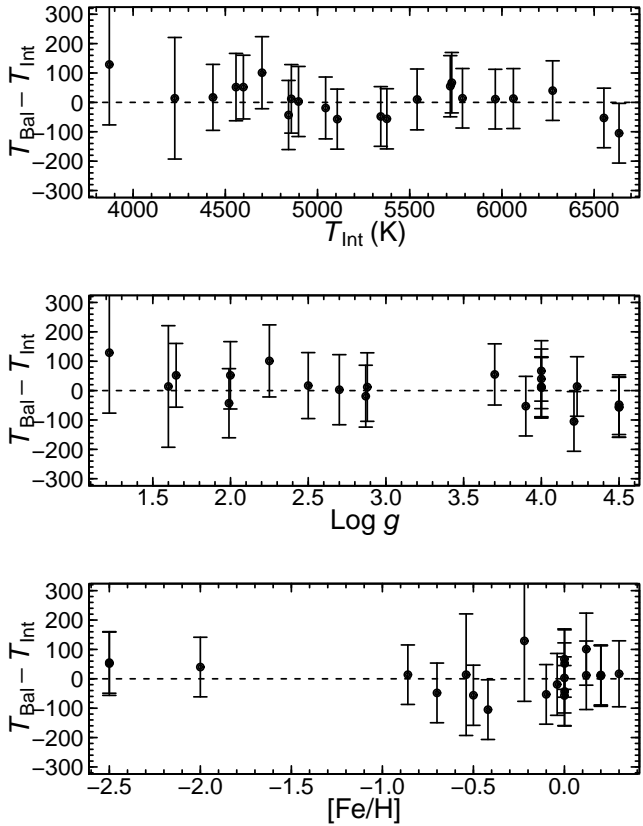}
\caption{Comparison of the Balmer effective temperature estimate to that from interferometry, as listed in Table~\ref{tab-inter}.  The Balmer effective temperature estimates agree with the interferometric estimate to within $\sim100$~K.}
\label{fig-int}
\end{figure}

Currently, the only way to understand whether our Balmer $\teff$ scale is affected by systematics is by comparing with independent methods, in particular interferometry. We, therefore, computed the Balmer $\teff$ for several nearby stars with direct and indirect interferometric angular diameter measurements. The results are listed in Table~\ref{tab-inter}, while we plot the difference between our Balmer estimate and that from interferometry in Figure~\ref{fig-int}. Both $\teff$ scales show an agreement of $3\pm60$~K for stars with $\feh > -1$, while the Balmer estimate is $\sim50$~K warmer than $\tint$ at the lowest metallicities.  These differences are well within the combined errors in the interferometric and Balmer measurements.  This suggests that deviations from 1D HE \textit{and} LTE are either minimal, or affect both interferometric and Balmer $\teff$ in exactly same way. Also note, for the stars in common with \citet{cayrel11}, our estimates are fully consistent.

\begin{table}
\centering
\begin{minipage}{80mm}
 \caption{Effective temperatures determined from direct interferometric measurements of angular diameters.}
 \label{tab-inter}
 \begin{tabular}{@{}rrrrrrrl@{}}
 \hline
  HD & $\logg$ & $\feh$ & $\tint$ & $\sigma_{\tint}$ & $\tbal$ & $\sigma_{\tbal}$ & Ref. \\
     &         &        & (K)  & (K)  & (K) & (K)  & \\
\hline
6582  & 4.50 & -0.70 & 5343 & 18 & 5295 & 100 & a\\
10700 & 4.50 & -0.50 & 5376 & 22 & 5320 & 100 & a\\
22049 & 4.50  & 0.00 & 5107 & 21 & 5050 & 100 & a\\
22879 & 4.23 & -0.86 & 5786 & 16 & 5800 & 100 & b*\\
27697 & 2.70  & 0.00 & 4897 & 65 & 4900 & 100 & c \\
28305 & 2.00  & 0.00 & 4843 & 62 & 4800 & 100 & c \\
29139 & 1.22 & -0.22 & 3871 & 48 & 4000 & 200 & c \\
49933 & 4.21 & -0.42 & 6635 & 18 & 6530 & 100 & b\\
61421 & 3.90 & -0.10 & 6555 & 17 & 6500 & 100 & a\\
62509 & 2.88  & 0.12 & 4858 & 60 & 4870 & 100 & c \\
84937 & 4.00 & -2.00 & 6275 & 17 & 6315 & 100 & b*\\
85503 & 2.50  & 0.30 & 4433 & 51 & 4450 & 100 & b\\
100407 & 2.87 & -0.04 & 5044 & 33 & 5025 & 100 & b\\
102870 & 4.00  & 0.20 & 6062 & 20 & 6075 & 100 & a\\
121370 & 4.00  & 0.20 & 5964 & 18 & 5975 & 100 & a\\
122563 & 1.65 & -2.50 & 4598 & 42 & 4650 & 100 & d\\
124897 & 1.60 & -0.54 & 4226 & 53 & 4240 & 200 & c \\
140283 & 3.70 & -2.50 & 5720 & 29 & 5775 & 100 & b*\\
140573 & 2.00  & 0.00 & 4558 & 56 & 4610 & 100 & c \\
150680 & 4.00  & 0.00 & 5728 & 24 & 5795 & 100 & a\\
161797 & 4.00  & 0.20 & 5540 & 27 & 5550 & 100 & a\\
215665 & 2.25  & 0.12 & 4699 & 71 & 4800 & 100 & c \\
\hline		
\end{tabular}
\end{minipage}
References for $\tint$: a - \citet{cayrel11}; b - {\it Gaia} calibration stars from U. Heiter (2012, priv. comm.) (Those marked with a [*] have angular diameters determined using the surface-brightness relations from \citealt{kervella04}); c - \citet{mozurkewich03}; d - \citet{creevey12}
\end{table}

For the second method, we utilised the Tycho-2 and 2MASS photometry of each star to compute effective temperature estimates, $\tirfm$, using the infrared flux method (IRFM), as presented in \citet[][hereafter C10]{casagrande10}.  Note that in C10 the IRFM calculations were applied only to dwarfs and
subgiants and validated by comparison with a large body of interferometric angular diameters. However, the same code can be safely applied to lower
surface gravities, as shown by comparison with newly determined angular diameters for giants \citep{huber12}. The advantage of this method is that it is much less sensitive to model assumptions that are required for spectroscopic analyses.  However, the quality of the photometric data used to compute IRFM effective temperatures, as well as interstellar reddening, can still largely affect the result.  In our case reddening has been estimated using the \citet{drimmel03} map, with distances derived from our spectroscopic $\logg$. Typical values of $E(B\mbox{-}V)$ are around 0.05~mag, although for some of the brightest giants the value can be considerably larger (see Table~\ref{tab-par}).

Finally, we chose the effective temperature estimates from R11 (denoted as $\tr11$) as our third effective temperature scale.  These estimates were based on the R11 calibration, which was derived from the the trend between $\fehlte$ and the difference between $\tlte$ and the 2MASS $\jks$ photometric effective temperature for several globular cluster stars, {\it Hipparcos} stars, and low-reddened stars in the R11 sample.  In principle, this method should yield similar effective temperature estimates to that of IRFM, since it utilises the colour-temperature transformations presented in \citet{ghernandez09}, which were based upon their IRFM calculations.  The advantage is that the R11 calibration relies on the $\jks$ colour, which is less sensitive to reddening ($E(\jks)\sim0.5E(B\mbox{-}V)$). However, note that the $\jks$ colour correlates only mildly with $\teff$, and thus calibrations involving that index exhibit a rather larger internal dispersion (in our case 139~K for dwarfs and 94~K for giants) when compared to IRFM effective temperature estimates for standard stars.  Further, effective temperatures computed using the calibration in C10 are typically $\sim40$~K hotter than those computed in \citet{ghernandez09}.  A detailed explanation for this discrepancy is given in C10.

\subsection{Comparisons}

We next applied each of the above methods to our sample stars, the values of which can be found in Table~\ref{tab-par}.  Note that for several stars, the Balmer lines fell in the middle of the order of the spectrum.  The continuum cannot be determined with sufficient accuracy in such regions.  We therefore did not measure the Balmer lines for those stars.  Further,  not all stars in our sample have Tycho-2 photometry estimates.  We were unable to compute an IRFM effective temperature estimate for these stars.

\begin{figure}
\includegraphics[width=84mm]{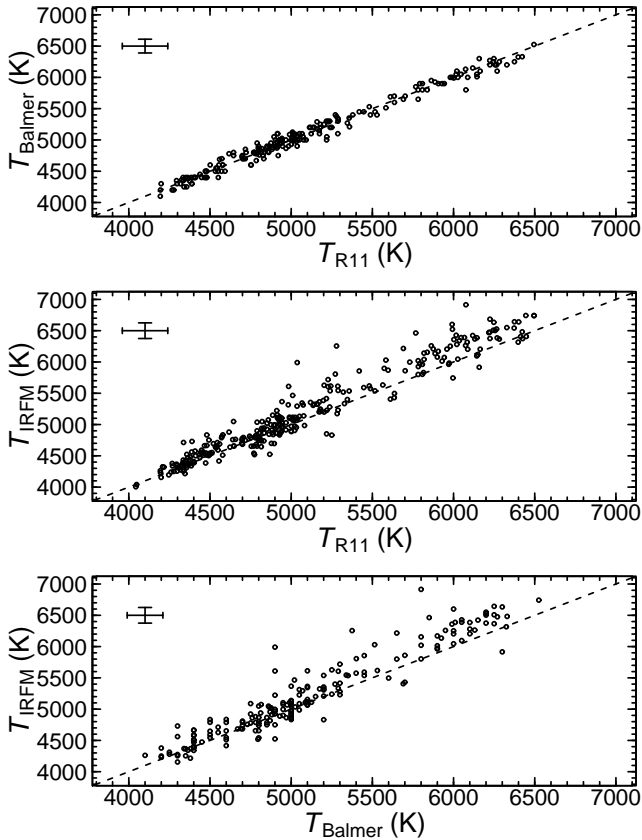}
\caption{Comparison of the three different $\teff$ scales.  Both the estimate from R11 ($\tr11$) and the Balmer lines ($\tbal$) agree within $\sim80$~K, while that from IRFM ($\tirfm$) is systematically higher than both, and shows an increasing dispersion with increasing effective temperature.  Error bars, which show the mean error in each $\teff$ scale, are displayed in the upper left corner.}
\label{fig-tcomp}
\end{figure}

Figure~\ref{fig-tcomp} shows the comparisons between the three effective temperature estimates when applied to our sample stars.  The estimates from $\tbal$ and $\tr11$ show remarkable agreement, with a difference of only $1\pm 79$~K.  The IRFM effective temperatures, however, are systematically higher than both $\tbal$ and $\tr11$ (by $119\pm215$~K and $113\pm186$~K, respectively), with an increasing dispersion towards hotter effective temperatures.  

It is possible that inherent NLTE and 3D effects could be influencing the Balmer effective temperature scale, however, we see excellent agreement with $\tr11$ and interferometric measurements.  As stated previously, the effective temperature estimates computed in C10 are about 40~K warmer than those computed in \citet{ghernandez09}.  Further, the $\jks$ calibration has a large internal dispersion.  However, the difference between $\tirfm$ and $\tr11$ extends well beyond these limits.  

It is possible that the uncertainty in the interstellar reddening may be systematically affecting the IRFM estimates.  In order to test the accuracy of the estimates of reddening from the \citet{drimmel03} map, we also tried to measure $E(B-V)$ using the interstellar \nion{Na}{D} lines.  However, the majority of the stars in our sample had multi-component interstellar \nion{Na}{D} features, or the feature was not discernible from the stellar Na lines.  Note, for several of the stars in which we could measure single-component interstellar \nion{Na}{D} lines, the Drimmel~et~al. estimates and the \nion{Na}{D} estimates were on average different by $<0.02$ mag, which translates to a difference of $<100$~K in effective temperature.  Given these differences, reddening alone cannot account for the very large differences ($\ge300$~K) for many stars.  

Instead, the large scatter mostly likely arises from the poor quality of the Tycho-2 magnitudes for stars fainter than $V_{\rm Tycho}\sim9$.  For the kind of stars analysed in this work, $B$ and $V$ magnitudes are the dominant contributors to the bolometric flux, as compared to the infrared 2MASS magnitudes.  Should a star be matched with a brighter (dimmer) source in $B$ and $V$, then the bolometric flux will be over-estimated (underestimated) by a very large amount, and IRFM will return a systematically higher (lower) $\tirfm$ estimate.  

\begin{figure}
\includegraphics[width=60mm,angle=-90]{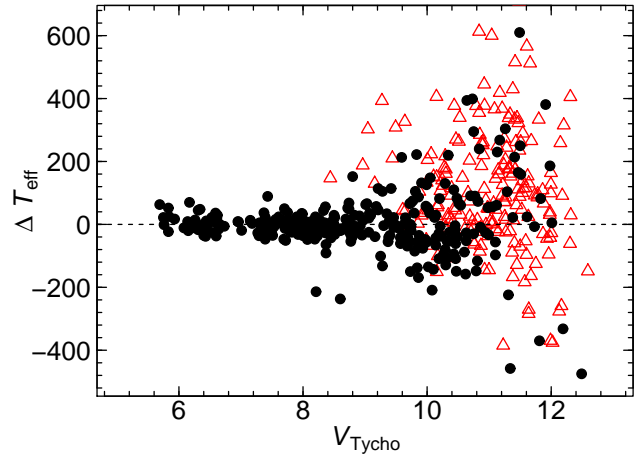}
\caption{An illustration of the difference between different effective temperature estimates as a function of $V_{\rm Tycho}$.  The black, filled circles represent the difference between $T(V_{\rm Tycho}\mbox{-}K_{\rm s})$ and $T(V_{\rm JC}\mbox{-}K_{\rm s})$ for the sample of stars in C10.  The red, open triangles show the difference between $\tirfm$ and $\tbal$ for the stars in our sample.  For $V_{\rm Tycho}\ge9$, both samples display an increased scatter in $\Delta\teff$.  This clearly indicates that the uncertainty in $V_{\rm Tycho}$ is poorly constrained for those stars.}
\label{fig-ty}
\end{figure}

Using the sample in C10, it is possible to compute $\teff$ using both the Tycho-2 $V_{\rm Tycho}\mbox{-}K_{\rm s}$ and Johnson-Cousins $V_{\rm JC}\mbox{-}K_{\rm s}$ calibrations.  We plot the difference between $T(V_{\rm Tycho}\mbox{-}K_{\rm s})$ and $T(V_{\rm JC}\mbox{-}K_{\rm s})$ as a function of the Tycho-2 $V$-magnitude for the C10 sample (black points) in Figure~\ref{fig-ty}.  In addition, we have over-plotted the difference between $\tirfm$ and $\tbal$ for the stars in our present sample (red triangles).  As shown in the figure, both samples exhibit a large scatter in the difference in effective temperature estimates at $V_{\rm Tycho}\ge9$.  Further, the hotter stars are on average among the faintest stars in our sample and thus have larger errors in $V_{\rm Tycho}$.  This is a clear indicator that the poor quality of the Tycho-2 photometric measurements for our stars is responsible for the discrepancy between $\tirfm$ and $\tbal$. 

\subsection{Final Effective Temperature Estimates}

From the comparisons above, Balmer line measurements provide the most reliable effective temperature estimates for all stars in our sample. In addition, $\tr11$ exhibits small differences with respect to $\tbal$ across all effective temperatures, as illustrated in Figure~\ref{fig-tcomp}.  In contrast, IRFM effective temperatures appear to show a large dispersion, which we attribute to large errors in Tycho-2 photometry, as well as uncertainty in the reddening.  We therefore adopted the mean of only $\tbal$ and $\tr11$, weighted according to the internal errors from each method, as our final $\tfin$ estimate.  This also serves to reduce the internal error on the final optimal effective temperature estimate, which was typically $\simlt100$~K.  

As noted previously, we could not measure the Balmer lines in several of the stars in our sample.  For those stars, we adopted the $\tr11$ estimate.  The authors of R11 adopted an error of 140~K in their $\tr11$ estimate, which was derived from the residuals in their calibration.  However, the comparison between $\tbal$ and $\tr11$ in Figure~\ref{fig-tcomp} suggests that this value was overestimated.  The mean error in $\tfin$, for those stars with both a $\tbal$ and $\tr11$ estimate, was 90~K.  We therefore adopted this value for stars with only a single $\tr11$ estimate.  Our final $\tfin$ values and corresponding errors can be found in Table~\ref{tab-par}.

\section{Surface Gravity and [Fe/H] in NLTE}
\label{sec-nlte}

Using the final values of $\tfin$ described in the previous section, we derived the remaining stellar parameters using the NLTE-Opt method described in Section~\ref{sec-par}.  We first validated this methodology by applying our analysis to a sample of 18 ``standard" stars, which have {\it Hipparcos} parallaxes.  The spectra for these stars were obtained for the analysis in \citet{fulbright00}, and are of similar quality to our sample. Both the LTE-Fe and NLTE-Opt atmospheric parameters for each standard star are given in Table~\ref{tab-hip}.  Using the {\it Hipparcos} parallax and an estimate of the bolometric correction derived from the bolometric flux relations presented in \citet{ghernandez09}, we also computed an ``astrometric surface gravity" ($g_{\pi}=4\pi GM\sigma T^4/L$) for each star, which is listed as $\loggpi$ in Table~\ref{tab-hip}.  Note, we computed an astrometric surface gravity using other various flux relations \citep{alonso95,casagrande10,torres10}, finding results within $\simlt0.1$~dex of that computed using the relation in \citet{ghernandez09}. Further, we assumed a mass of $0.8~M_{\odot}$ for $\feh<-1$ and $0.9~M_{\odot}$ for $\feh\ge-1$.  However, a difference of $0.1~M_{\odot}$ will only change the astrometric surface gravity by $\sim0.05$~dex. The NLTE-Opt surface gravity estimates show a remarkable $0.02\pm0.11$~dex agreement with the astrometric surface gravity, while the LTE-Fe estimates are too low by $-0.32\pm0.39$~dex.

Given the agreement, we applied the above analysis to our sample stars.  The final NLTE-Opt estimates for surface gravity and metallicity, as well as for the microturbulence, can be found in Table~\ref{tab-par}.  We adopted 0.1~dex error in both the surface gravity and metallicity, based on our comparisons with the standard stars above.

\begin{table*}
\centering
 \caption{{\it Hipparcos} Star Surface Gravity Comparisons.}
 \label{tab-hip}
  \begin{tabular}{@{}rrrrrrrr@{}}
  \hline
   HD & $\tlte$ & $\logglte$ & $\fehlte$ & $\tfin$ & $\loggf$ & $\fehf$ & $\loggpi$ \\
     {\it err}    &     ($\sim\pm60$~K)    &    ($\pm0.1$)    &  ($\pm0.1$) & ($<\pm100$~K)  & ($\pm0.1$) & ($\pm0.1$) & ($\pm0.1$) \\
   \hline
22879 & 5726 & 4.04 & -0.92 & 5817 & 4.27 & -0.89 & 4.33 \\
24616 & 5084 & 3.34 & -0.62 & 5071 & 3.40 & -0.69 & 3.29 \\
59374 & 5741 & 4.04 & -0.96 & 5877 & 4.33 & -0.88 & 4.49 \\
84937 & 6137 & 3.58 & -2.34 & 6374 & 4.18 & -2.11 & 4.15 \\
108317 & 4922 & 1.89 & -2.58 & 5367 & 3.04 & -2.14 & 3.14 \\
111721 & 4956 & 2.52 & -1.37 & 5091 & 2.93 & -1.29 & 2.70 \\
122956 & 4569 & 1.15 & -1.75 & 4750 & 1.94 & -1.61 & 2.03 \\
134169 & 5868 & 4.03 & -0.77 & 5924 & 4.20 & -0.74 & 4.03 \\
140283 & 5413 & 2.81 & -2.79 & 5834 & 3.71 & -2.41 & 3.73 \\
157466 & 6070 & 4.41 & -0.34 & 6002 & 4.37 & -0.41 & 4.35 \\
160693 & 5808 & 4.29 & -0.47 & 5749 & 4.24 & -0.55 & 4.31 \\
184499 & 5740 & 4.11 & -0.58 & 5766 & 4.23 & -0.57 & 4.08 \\
193901 & 5555 & 3.94 & -1.18 & 5775 & 4.39 & -1.01 & 4.57 \\
194598 & 5814 & 4.02 & -1.23 & 5991 & 4.39 & -1.10 & 4.27 \\
201891 & 5676 & 3.89 & -1.21 & 5871 & 4.30 & -1.06 & 4.30 \\
204155 & 5696 & 3.94 & -0.71 & 5733 & 4.08 & -0.69 & 4.03 \\
207978 & 6343 & 3.93 & -0.62 & 6294 & 4.02 & -0.62 & 3.96 \\
222794 & 5588 & 3.99 & -0.66 & 5604 & 4.08 & -0.66 & 3.91 \\
\hline		
\end{tabular}
\end{table*}

\section{NLTE-Opt vs. LTE-Fe}

\begin{figure}
\includegraphics[width=84mm]{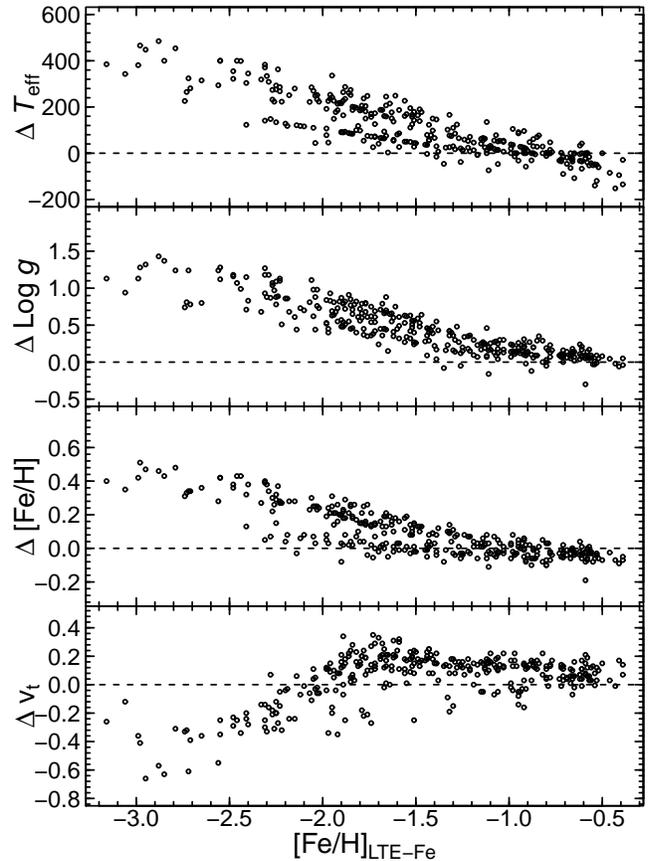}
\caption{Comparison of stellar parameters derived using the LTE-Fe method and the NLTE-Opt stellar parameters vs. $\fehlte$.  The difference in effective temperature, surface gravity, and metallicity show a large systematic increase with decreasing metallicity. The dual trends seen in $\Delta~\teff$, $\Delta~\logg$, and $\Delta~\feh$ are a result of the R11 effective temperature calibration, in which the authors found that stars with effective temperatures less than 4500~K only required a small correction to $\tlte$.  Therefore, these stars stand out in the plots.}
\label{fig-lte}
\end{figure}

In Figure~\ref{fig-lte}, we compare our final NLTE-Opt stellar parameters to those derived using the LTE-Fe method. The differences in the estimates of  effective temperature, surface gravity, metallicity, and microturbulence all display clear trends with decreasing metallicity. The microturbulent velocity is underestimated by $\sim0.1\mbox{-}0.2$~\kmsec{} until $\fehlte\sim-1.8$, where $\vtlte$ becomes larger than $\vtf$. The differences between $\teff$ range from $200$ to $400$ K for metal-poor giants, and $-50$ to $-100$ K for dwarfs. The differences for $\log g$ and [Fe/H] reach a factor of $30$ in surface gravity ($\Delta \log g = 1.5$ dex) and a factor of $3$ in metallicity ($\Delta$ [Fe/H] $=0.5$ dex) at [Fe/H] $\sim -2.5$.

\begin{figure}
\includegraphics[width=51mm,angle=-90]{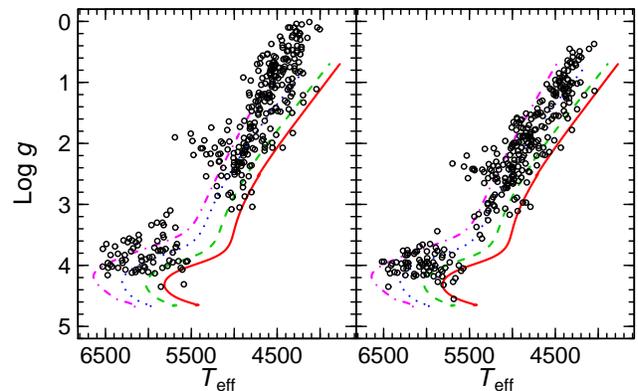}
\caption{Stellar positions in the $\logg$ vs. $\teff$ plane in LTE-Fe (left panel) and NLTE-Opt (right panel).  The curves shown are evolutionary tracks computed using the GARSTEC code \citep{weiss08}.  Each track was computed assuming a mass of $0.8~M_{\odot}$ and an $\feh$ of -0.5 (solid, red), -1.0 (long-dashed, green), -1.5 (short-dashed, blue), and -2.5 (dot-dashed, magenta).  Note, those stars around $\logg=2.0$ that lie away from the tracks are most likely horizontal branch stars.}
\label{fig-hrd}
\end{figure}

Figure~\ref{fig-hrd} illustrates how the different LTE-Fe and NLTE-Opt results can change the position of each star in the $\logg$ vs. $\log(\teff)$ plane.  In addition, we have included several evolutionary tracks, computed using the GARSTEC code \citep{weiss08}, for comparison.  Generally, the NLTE-Opt estimates of surface gravity and effective temperature trace the morphology of the theoretical tracks much more accurately.  Several features are most notable.  The NLTE-Opt parameters lead to far less stars that lie on or above the tip of the red giant branch, and more stars occupy the middle or lower portion of the RGB. Also, stars at the turn-off and subgiant branch are now more consistent with stellar evolution calculations.

\begin{figure}
\includegraphics[width=68mm,angle=-90]{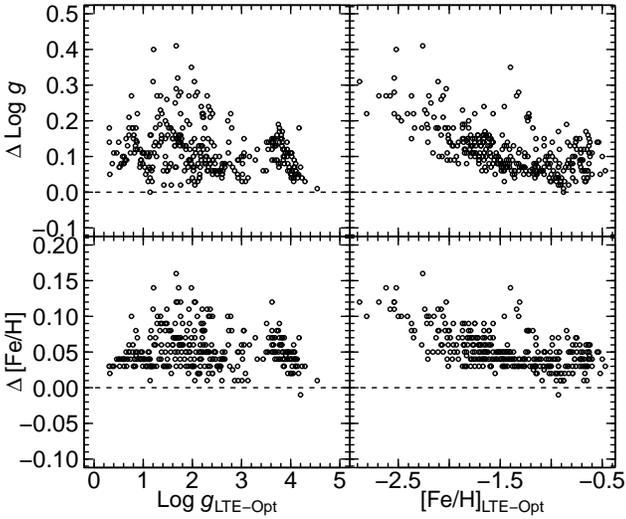}
\caption{Comparison of LTE-Opt parameters to NLTE-Opt parameters.  This shows the underlying effects of the NLTE corrections.  Typically, the LTE-Opt surface gravity and metallicity estimates are systematic underestimated by about 0.1~dex and 0.05~dex, respectively.  However, they become increasingly more underestimated to about 0.15~dex and 0.3~dex, respectively, at the lowest metallicities.}
\label{fig-copt}
\end{figure}

Figures~\ref{fig-lte} and \ref{fig-hrd} further prompted us to determine the relative importance of the effective temperature scale versus the NLTE corrections for gravities and metallicities in the NLTE-Opt method.  We singled out the effect of the NLTE corrections by deriving additional, LTE-Opt surface gravity and metallicity estimates using LTE iron abundances combined with our $\tfin$ estimate.  Note, as with the NLTE-Opt method, Fe lines which have an excitation potential below 2~eV were excluded.  The comparison between these LTE-Opt estimates and the final NLTE-Opt estimates are shown in Figure~\ref{fig-copt}.  As evident from this figure, solving for ionisation equilibrium in NLTE also leads to \textit{systematic} changes in the $\log g$ and [Fe/H], such that LTE gravities are under-estimated by $0.1$ to $0.3$~dex, whereas the error in metallicity is about $0.05$ to $0.15$~dex.  These effects are consistent with that seen in \citet{lind12}.

We thus conclude that reliable effective temperatures are necessary to avoid substantial biases in a spectroscopic determination of $\log g$ and [Fe/H], such as displayed in Figure~\ref{fig-lte}.  We have shown here that, at present, excitation balance of \nion{Fe}{I} lines with 1D hydrostatic model atmospheres in LTE does not provide the correct effective temperature scale, supporting the results by Bergemann et al. (2012).  On the contrary, Balmer lines provide such a scale. Furthermore, NLTE effects on ionisation balance are necessary to eliminate the discrepancy between \nion{Fe}{I} and \nion{Fe}{II} lines, an effect that is present, regardless of the adopted $\teff$. Only in this way is it possible to determine accurate surface gravity and metallicity from Fe lines.

\section{Conclusion}

In this work, we explore several available methods to determine effective temperature, surface gravity, and metallicity for late-type stars. The methods include excitation and ionization balance of Fe lines in LTE and NLTE, semi-empirically calibrated photometry (R11), and the Infra-Red flux method (IRFM). Applying these methods to the large set of high-resolution spectra of metal-poor FGK stars selected from the RAVE survey, we then devise a new efficient strategy which provides robust estimates of their atmospheric parameters. The principal components of our method are (i) Balmer lines to determine effective temperatures, (ii) NLTE ionization balance of Fe to determine $\logg$ and $\feh$, and (iii) restriction of the \nion{Fe}{I} lines to that with the lower level excitation potential greater than 2~eV to minimize the influence of 3D effects \citep{bergemann12}.

A comparison of the new NLTE-Opt stellar parameters to that obtained from the widely-used method of 1D LTE excitation-ionization of Fe, LTE-Fe, reveals significant \textit{systematic biases} in the latter. The difference between the NLTE-Opt and LTE-Fe parameters systematically increase with decreasing metallicity, and can be quite large for the metal-poor stars: from 200 to 400~K in $\teff$ , 0.5 to 1.5~dex in $\logg$, and 0.1 to 0.5~dex in $\feh$.  These systematic trends are largely influenced by the difference in the estimate of the stellar effective temperature, and thus, a reliable effective temperature scale, such as the Balmer scale, is of critical importance in any spectral parameter analysis.  However, a disparity between the abundance of iron from \nion{Fe}{I} and \nion{Fe}{II} lines still remains. It is therefore necessary to include the NLTE effects in \nion{Fe}{I} lines to eliminate this discrepancy. 

The implications of the very large differences between the NLTE-Opt and LTE-Fe estimates of atmospheric parameters extend beyond that of just the characterisation of stars by their surface parameters and abundance analyses. Spectroscopically derived parameters are often used to derive other fundamental stellar parameters such as mass, age and distance through comparison to stellar evolution models.  The placement of a star along a given model will be largely influenced by the method used to determine the stellar parameters.  For example, distance scales will change, which could affect the abundance gradients measured in the Milky Way (e.g., R11), as well as the controversial identification of different components in the MW halo \citep{schonrich11,beers12}.  We explore this in greater detail in the next paper of this series \citep{serenelli12}.

\section*{Acknowledgements}

We acknowledge valuable discussions with Martin Asplund, and are indebted to Ulrike Heiter for kindly providing interferometric temperatures for several {\it Gaia} calibration stars.  We also acknowledge the staff members of Siding Spring Observatory, La Silla Observatory, 
Apache Point Observatory, and Las Campanas Observatory for their assistance in making the observations for this project possible.  Greg Ruchti acknowledges support through grants from ESF EuroGenesis and Max Planck Society for the FirstStars collaboration.  Aldo Serenelli is partially supported  by the  European Union International Reintegration Grant PIRG-GA-2009-247732, the MICINN grant AYA2011-24704, by the ESF EUROCORES Program EuroGENESIS (MICINN grant EUI2009-04170), by SGR grants of the Generalitat de Catalunya and by the EU-FEDER funds.

\clearpage

\end{document}